\documentclass{aa}
\usepackage{graphicx}
\usepackage{txfonts}
\newcommand{\re}{R_{\rm Earth}}
\newcommand{\rs}{R_{\rm Sat}}
\newcommand{\zmax}{z_{\rm max}}
\newcommand{\dg}{^{\circ}}

\newcommand{\B}[1]{#1}
\usepackage{natbib}
\bibpunct{(}{)}{;}{a}{}{,} 

\begin{document}

\title{Impact of satellite constellations on astronomical observations with ESO telescopes in the visible and infrared domains}
\author{Olivier R. Hainaut \inst{\ref{inst:eso}}\and
Andrew P. Williams \inst{\ref{inst:eso}}
}
\institute{
    {European Southern Observatory,
    Karl-Schwarzschild-Strasse 2,
    D-85748 Garching bei M\"unchen, Germany}\label{inst:eso}
}
\date{Submitted 2020-Jan-14 AA/2020/37501 -- Accepted 2020-Feb-28
}

\abstract{    
The effect of satellite mega-constellations on astronomical observations in the visible, near-infrared, and thermal infrared domains is estimated using a simple methodology, which is applied to ESO telescopes and instruments as examples (radio and (sub-)millimetre domains are not considered here). The study considers a total of 18 constellations in development by SpaceX, Amazon, OneWeb, and others, with over 26 thousand satellites, constituting a representative distribution. This study uses a series of simplifications and assumptions in order to obtain conservative, order-of-magnitude estimates of the effects: the satellites are assumed to be uniformly spread over the Earth's globe, and their magnitude is estimated using a simplistic model calibrated on actual observations. The effect on various types of ground-based telescopic observations is estimated using a geometric probabilistic approach.

The `trains' of very-low altitude satellites typically observed immediately after launch are extremely bright due to their very low orbit. They also fall very quickly in the shadow of the Earth after sunset. However, this initial bright state is not considered further, as the satellites quickly disperse into their higher altitude orbits.

The number of illuminated satellites {from the constellations} above the horizon of an observatory ranges from approximately 1600 immediately after sunset, decreasing to 1100 at the end of astronomical twilight, most of them ($\sim 85$\%) close to the horizon (below 30$\dg$ of elevation). The large majority of these satellites will be too faint to be seen with the naked eye:
at astronomical twilight, 260 would be brighter than magnitude 6 (i.e. visible in exceptional conditions), 110 brighter than 5 (i.e. visible in good conditions). Again, most of them ($\sim 95$\%) will be close to the horizon (below 30$\dg$ of elevation). The number of naked-eye satellites plummets as the Sun reaches 30--40$\dg$ of elevation below the horizon.

Specular flares and occultations by satellites are expected to cause only negligible effects on telescopic astronomical observations. The light trail caused by the satellite would ruin a small fraction (below the 1\% level) of telescopic exposures using narrow to normal field imaging or spectroscopic techniques in the visible and near-infrared during the first and last hours of the night. Similarly, the thermal emission of the satellite would affect only a negligible fraction of the observations in the thermal infrared domain. 
However, wide-field exposures and long medium-field exposures would be affected at the 3\% level during the first and last hours of the night. Furthermore, ultra-wide imaging exposures on a very large telescope (where saturation of the satellite trails has a ruinous effect on the detectors, such as those from the National Science Foundation's Vera C. Rubin Observatory, formerly known as LSST), would be significantly affected, with 30 to 40\% of such exposures being compromised during the first and last hours of the night. 
Coordination and collaboration between the astronomical community, satellite companies, and government agencies are therefore critical to minimise and mitigate the effect on astronomical observations, in particular on survey telescopes. 
 }

\keywords{Astronomical instrumentation, methods and techniques; Light pollution; Methods: observational; Site testing}
          
\titlerunning{Satellite Constellations}
\authorrunning{Hainaut O.~R.}

\maketitle
\section{Introduction}
\B{While artificial satellites have until now been a substantial concern for radio astronomy and a relatively minor issue for observers in the optical spectrum, the recent launches of the SpaceX Starlink constellation with their spectacularly bright post-launch appearance, and the growing publicity of the plans of other companies for major constellations of thousands of satellites, have caused alarm in the astronomy community. The issue has also generated substantial media attention, which has highlighted the emotional and moral dimensions of the issue that go beyond the impacts on astronomical science. In order for the astronomy community to respond to these developments and work constructively with industry, funding agencies, and regulators, a factual and quantitative assessment of the impacts is required.}

This paper aims at quantifying the effect of large satellite constellations on visible, near-infrared (NIR) and thermal IR astronomical observations using a series of simplifying assumptions. It does not replace careful, detailed simulations taking into account the intricacies of the orbital distribution, the complexity of estimating the brightness of a satellite, and so on, but provides a first, quantitative estimate of the effect. As the simplifications and assumptions used are conservative, the estimated effect is likely greater than the actual effect.

A series of known constellations were taken into account, totalling 18 (sub-) constellations and over 26\,000 satellites. A simple approximation was used to model their distribution. 

The effect on visible and near-infrared observations (NIR) was estimated using a simple model for their brightness, which computes their magnitude as a function of their orbital altitude and of the angular elevation above the horizon. While this model is extremely crude, it is calibrated using known satellites and observations of SpaceX's recently launched `Starlink' satellites, and is validated by direct photometric observations of a Starlink satellite. Additionally, an estimate of the number of specular reflections -- the bright satellite flares -- is provided by scaling the flares observed for the Iridium first-generation satellites.
Accounting for all of the above, the effect is computed on various types of observations: different exposure duration, various field-of-view sizes, visible and thermal-infrared observations, and also occultation by a non-illuminated satellite passing in front of the object observed. Computation was performed using a geometric probabilistic approach --what fraction of the sky would be contaminated by satellites for the considered observation type-- rather than a direct simulation involving repeated modelled observations. 

The effects on millimetre and submillimetre observations are not considered here. A separate paper will estimate these effects. \B{This paper focuses on the effect on pointed observations; other science cases, in particular wide-field observations such as surveys, could be more affected. Also not considered are the impacts on the amateur astronomy and astrophotography community.} The effects on observatory operations, associated cost implications, and political or regulatory issues are beyond the scope of this paper. 


\section{Constellation and number of satellites}
\subsection{Known upcoming constellations with orbits}

Table~\ref{tab:const} lists publicly known future mega-constellations that are in development, the number of satellites that are planned for launch, and the orbital altitude $h$ of the satellites. \B{The list is neither complete nor accurate, as it is based on generic web searches, operator websites, and on official documents submitted to the Federal Communications Commission (FCC).
Some of the constellations have already been cancelled. Other, non-telecommunication constellations have filed documentation with the FCC. Overall, this list should therefore be considered only as a representative list of a variety of constellations and a large number of satellites, rather than an exact representation of what will be launched in the coming years, which is sufficient for this study. Its results can be scaled to smaller or larger constellations.} The original Iridium constellation is also included, although its size and effects are negligible compared to the others. It is used for calibration of the flare numbers.
\begin{table*}
\begin{tabular}{lrr | r | rrr | rrr | rrr }
Constellation &  &   & $a_\sun$   & \multicolumn{3}{c|}{Satellites}    &\multicolumn{3}{c|}{Satellites}            &       &               &                \\
              &  &Alt.& for   & \multicolumn{3}{c|}{above horizon} &\multicolumn{3}{c|}{above $\zmax=60\dg$}  &Orbital&\multicolumn{2}{c}{Magnitude at}\\
              &  &    & Zenith   &          &        &                &           &           &                   & Period& Zenith        & $z=60\dg$ \\
             &$N$& $h$&$\gamma_o$ &$a_\sun$ & \% & $N$ & $a_\sun$& \% & $N$ & $P$   & &  \\
             &   &[km]&[deg]    &  [deg]     &    &     &[deg]     &    &     &[h]    &[mag]      &[mag]\\
& \footnotesize 1& \footnotesize  2& \footnotesize  3& \footnotesize  4& \footnotesize  5& \footnotesize  6& \footnotesize  7& \footnotesize  8& \footnotesize  9& \footnotesize  10& \footnotesize  11& \footnotesize  12\\
\hline
\hline
SpaceX Starlink \\
  ~~ 340                & 7518 & 340    &-18.3& -36.6   & 2.5\% & 190.3 & -23.0   & 0.2\% & 12.6  & 1.51  & 3.2   & 4.8\\
~~ 550                  & 1600 & 550    &-23.0& -46.0   & 4.0\% & 63.5  & -30.1   & 0.4\% & 6.2   & 1.58  & 4.2   & 5.9\\
~~ 1150                 & 2800 & 1150   &-32.1& -64.2   & 7.6\% & 214.0 & -44.9   & 1.2\% & 34.8  & 1.79  & 5.8   & 7.5\\
OneWeb                  & 648  & 1200   &-32.7& -65.4   & 7.9\% & 51.3  & -45.9   & 1.3\% & 8.6   & 1.81  & 5.9   & 7.5\\
Amazon Kuiper\\                                
~~590                   & 784  & 590    &-23.8& -47.5   & 4.2\% & 33.2  & -31.3   & 0.4\% & 3.4   & 1.60  & 4.4   & 6.0\\
~~610                   & 1296 & 610    &-24.1& -48.2   & 4.4\% & 56.6  & -31.9   & 0.5\% & 6.0   & 1.60  & 4.5   & 6.1\\
~~630                   & 1156 & 630    &-24.5& -49.0   & 4.5\% & 52.0  & -32.5   & 0.5\% & 5.6   & 1.61  & 4.5   & 6.1\\
Sat Revolution          & 1024 & 350    &-18.6& -37.1   & 2.6\% & 26.6  & -23.4   & 0.2\% & 1.8   & 1.52  & 3.2   & 4.9\\
China CASC              & 320  & 1100   &-31.5& -63.0   & 7.4\% & 23.5  & -43.9   & 1.2\% & 3.7   & 1.78  & 5.7   & 7.4\\
China LuckyStar        & 156  & 1000    &-30.2& -60.4   & 6.8\% & 10.6  & -41.7   & 1.0\% & 1.6   & 1.74  & 5.5   & 7.2\\
China Commsat           & 800  & 600    &-23.9& -47.9   & 4.3\% & 34.4  & -31.6   & 0.4\% & 3.6   & 1.60  & 4.4   & 6.0\\
China Xinwei            & 32   & 600    &-23.9& -47.9   & 4.3\% & 1.4   & -31.6   & 0.4\% & 0.1   & 1.60  & 4.4   & 6.0\\
India AstroTech       & 600  & 1400     &-34.9& -69.8   & 9.0\% & 54.0  & -49.7   & 1.6\% & 9.9   & 1.88  & 6.3   & 7.9\\
Boing                   & 2956 & 1030   &-30.6& -61.2   & 7.0\% & 205.6 & -42.4   & 1.1\% & 31.2  & 1.75  & 5.6   & 7.2\\
LeoSat                  & 108  & 1423   &-35.2& -70.3   & 9.1\% & 9.9   & -50.1   & 1.7\% & 1.8   & 1.89  & 6.3   & 7.9\\
Samsung                 & 4700 & 2000   &-40.4& -80.9   & 11.9\%& 561.2 & -59.2   & 2.7\% & 124.8 & 2.11  & 7.0   & 8.7\\
Yaliny                  & 135  & 600    &-23.9& -47.9   & 4.3\% & 5.8   & -31.6   & 0.4\% & 0.6   & 1.60  & 4.4   & 6.0\\
Telesat LEO             & 117  & 1000   &-30.2& -60.4   & 6.8\% & 7.9   & -41.7   & 1.0\% & 1.2   & 1.74  & 5.5   & 7.2\\
\hline                                                                                 
(Iridium)               & 66   & 780    &-27.0& -54.0   & 5.5\% & 3.6   & -36.5   & 0.7\% & 0.5   & 1.66  & 5.0   & 6.6\\
\hline                                                                                          
Total                   & 26,750&        &    &   &       & 1,605 &       &       &       258 &&&\\                              
\end{tabular}

    \caption{List of the constellations used in this study; it is meant to provide a representative sample. The list includes the number of satellites (column 1) and their altitude in km (column 2). Column 3 is the elevation of the Sun for which a satellite at Zenith is just illuminated, also for which half the satellites in range are illuminated; by construction, it is also the angle Zenith-Centre of Earth-Satellite at which the satellite is on the horizon. 
    Considering the cases of satellites above the horizon and above a zenithal distance of $60\dg$, the table also lists the minimum solar elevation required to illuminate all these satellites  (Columns 4 and 7); the fraction of the constellation that is in range (Columns 5 and 8) and the corresponding number of satellites in range (columns 6 9). For reference, the orbital period is listed (column 10), as well as the magnitude of a satellite at zenith (column 11) and at $z=60\dg$ (column 12), for an intermediate solar phase angle of $90\dg$, considering our simple photometric model. 
    }
    \label{tab:const}
\end{table*} 

\begin{figure}
    \includegraphics[width=0.45\textwidth]{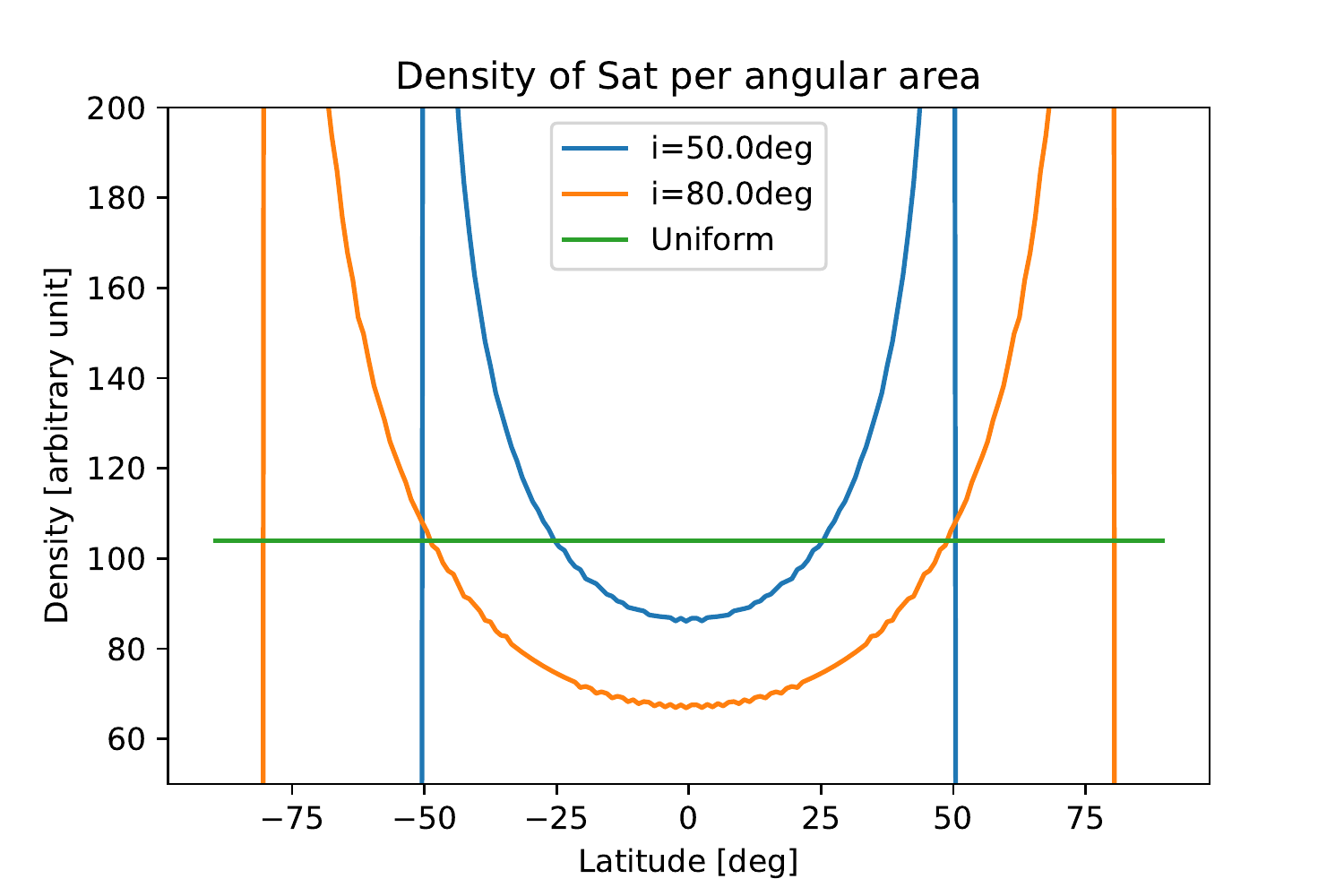}
    \caption{Density of satellites on their orbital sphere as a function of latitude, for two Walker constellations with orbital inclinations of 50 and 80$\dg$ and for the uniform approximation used in this paper. The total number of satellites is the same in all cases.}
    \label{fig:satDensity}
\end{figure}

\begin{figure*}
    \includegraphics[width=0.95\textwidth]{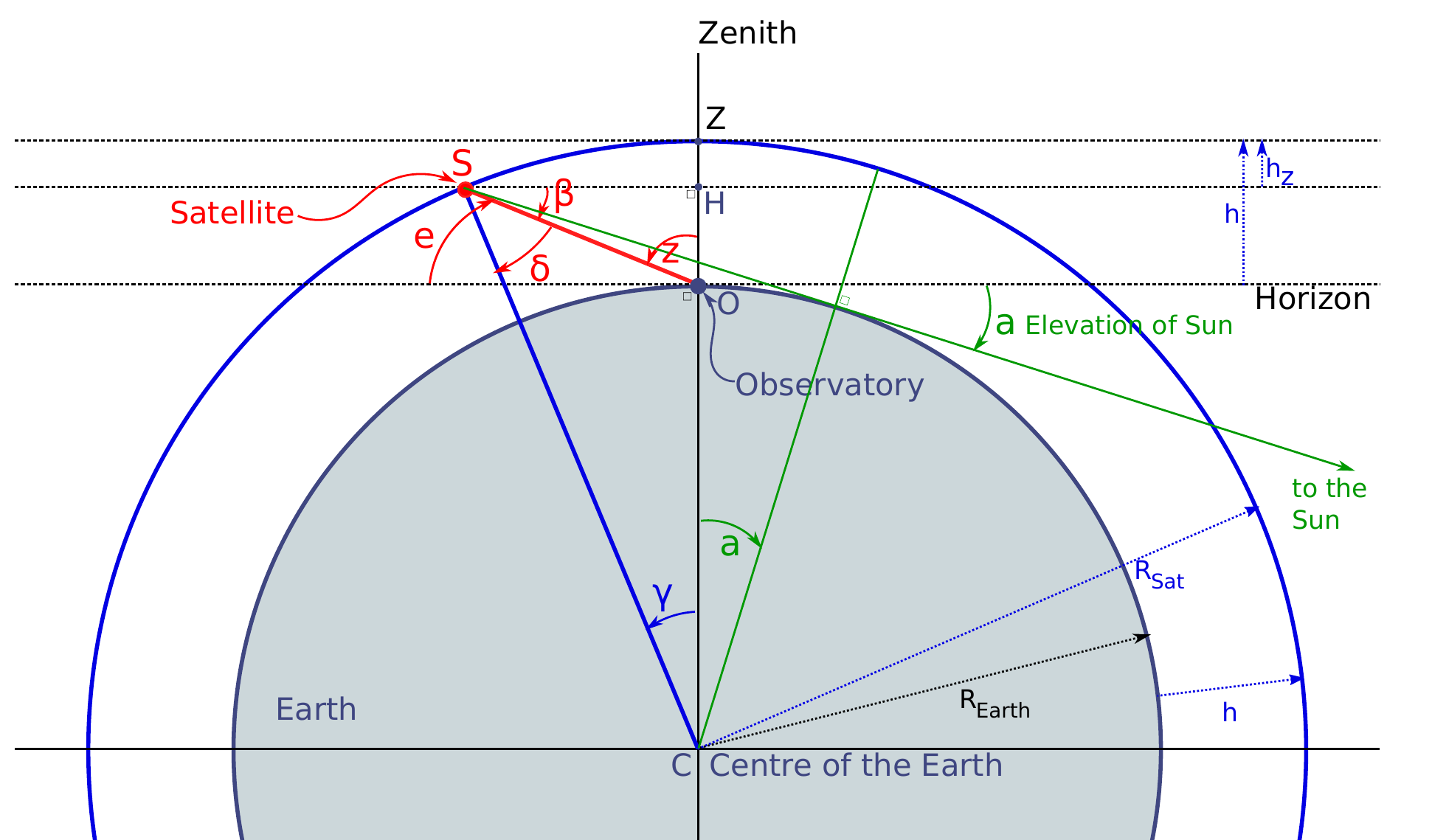}
    \caption{Angle and vector definitions. In this figure, the altitude of the satellite is $h \sim 1000$~km and the Sun is at the lowest elevation that illuminates the satellite.}
    \label{fig:satAngles}
\end{figure*}

\begin{figure}
    \centering
    a \includegraphics[width=0.4\textwidth]{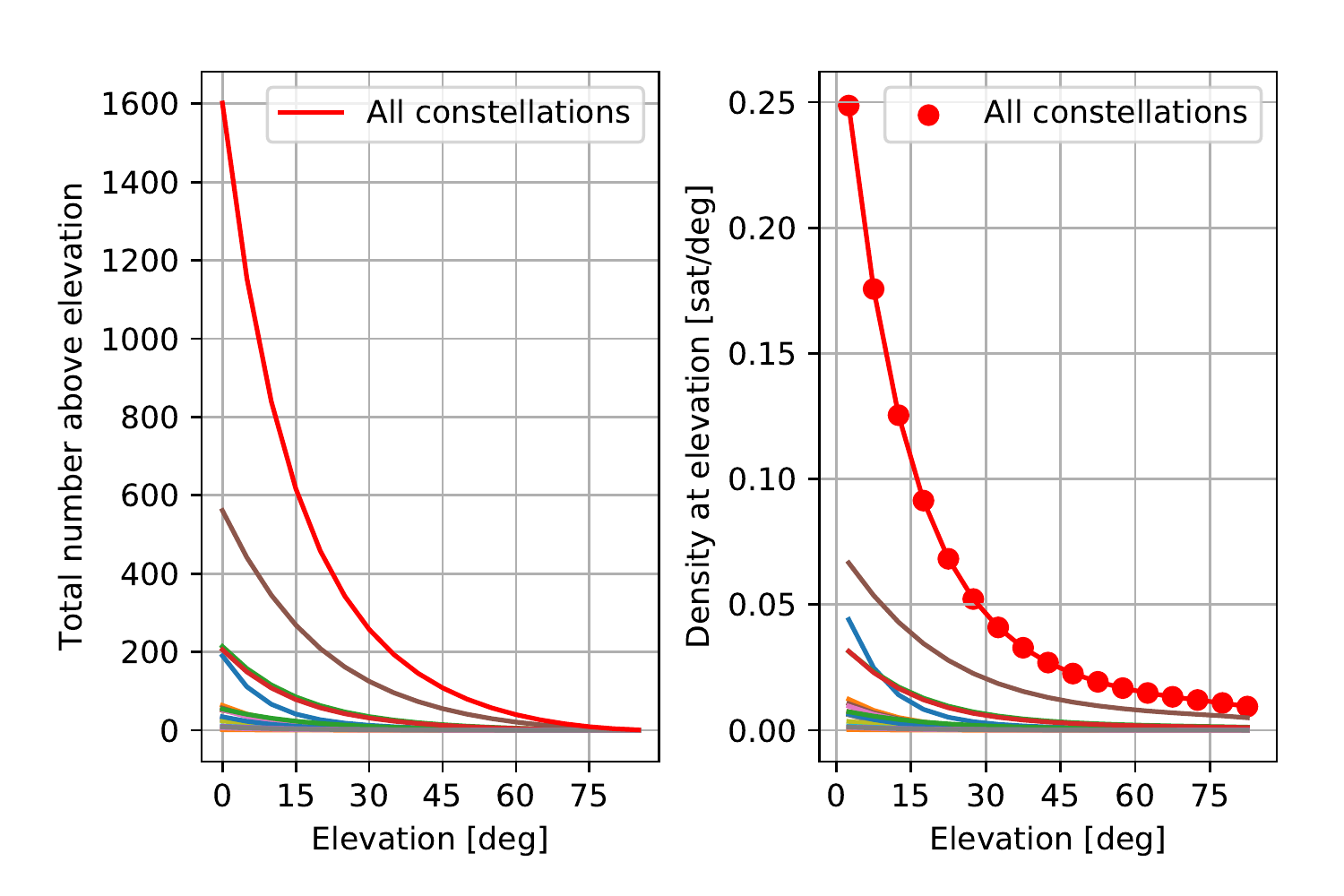}\\
    ~\\
    b \includegraphics[width=0.4\textwidth]{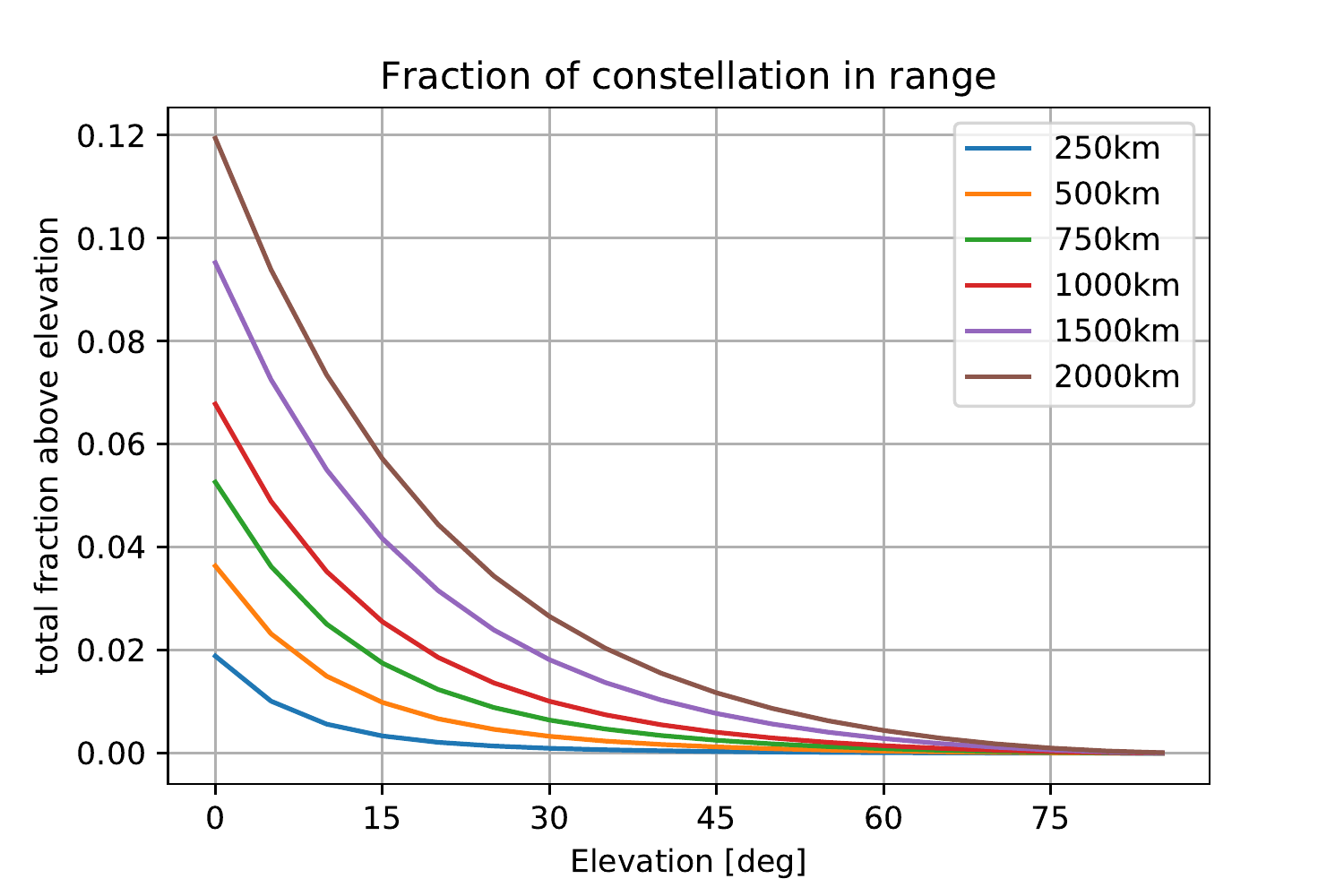}
    \caption{{\bf (a)} Left: Number of satellites above a given elevation; Right: Corresponding density (in satellite per square degree) at a given elevation, for all the constellations considered. The total value is in red. {\bf (b)}: Fraction of a constellation above a given elevation for a series of altitudes.
    }
    \label{Fig:fractions}
\end{figure}

\subsection{Distribution of the satellites}

The satellites in these mega-constellations will likely be organised in configurations similar to the Walker constellation \citep{walker84}, that is, all satellites from a constellation are on similar circular orbits with the same altitude and the same high inclination, grouped on a series of orbital planes whose nodes are uniformly distributed along the equator. The inclinations of the constellations considered are in the range of 42 to 80$\dg$ . However, in what follows,  the actual distribution is simplified: the satellites are assumed to be uniformly distributed over the whole Earth.  The actual Walker constellation distribution causes a dearth of satellites in the polar regions and an increase of the number of satellites at latitudes close to the inclination of the constellation. This approximation will therefore overestimate the number of satellites above the equatorial and low-latitude regions, will underestimate this number at  latitudes close to the orbital inclination, and will overestimate again the number in regions  above very high-latitudes, as illustrated in Fig.~\ref{fig:satDensity}. Many of the large professional telescopes are at latitudes lower than $\sim 30\dg$; for them, this approximation is conservatively overestimating the number of satellites. In this simplified configuration, the latitude of an observatory has no effect on the number of satellites that affect it.

\subsection{Number of satellites in range}
We now estimate the number of satellites above the horizon of an observatory (zenithal distance $z=90\dg$), and above $z=60\dg$ (elevation $e=60\dg$ above the horizon, also corresponding to an airmass of 2, the limit above which most of the astronomical observations are performed). At this stage, we consider only whether the satellite is in range, independently of it being illuminated or not. 

To do this, we compute the area of the spherical cap above the threshold value of $z$; the number of satellites present in that cap is then approximated by the ratio of the cap area to the area of the sphere, multiplied by the number of satellites on the sphere.

Here, we introduce $\gamma$, the orbital position angle, measured between the satellite and the zenith measured at the centre of the Earth. The condition of visibility of a satellite above the horizon, $z < 90\dg$, converts into
$\gamma< \gamma_o$, with 
\begin{equation}\label{Eq:gamma_o}
    \gamma_o = \arccos( \re / \rs) = \arccos \left( \frac{\re }{\re+h} \right),
\end{equation}
where $r_{\rm Earth} = 6375$~km is the radius of the Earth, $r_{\rm Sat} = r_{\rm Earth} + h $ is the radius of the satellite's orbit, and $h$ the altitude of the satellite above the Earth. 
The value of $\gamma_o$ is reported in Table~\ref{tab:const} for the various constellations. 
To generalise Eq.~\ref{Eq:gamma_o} to any value of $z$, we first introduce the angles $\beta = \pi/2 - z$ (angle O$\hat{\rm S}$H, opposite to $z$ in the right triangle OHS; see Fig.~\ref{fig:satAngles}) and $\delta$ (angle O$\hat{\rm S}$C). From the sinus theorem applied to triangle OCS,
\begin{equation}
    \frac{\sin{\delta}}{\re} = \frac{\sin{(\pi-z)}}{\rs} ~,
\end{equation}
or 
\begin{equation}\label{Eq:delta}
    \sin{\delta} = \frac{\re \sin{z}}{\rs} ~.   
\end{equation}
Summing the angles of the triangle OSC, 
\begin{equation}\label{Eq:triangle}
    \pi = \gamma + \delta + (\pi -z) ~.
\end{equation}
Using Equations \ref{Eq:gamma_o} and \ref{Eq:triangle},
\begin{equation}\label{Eq:gamma}
    \gamma = z - \arcsin{\left(\frac{\re}{\rs} \sin{z}\right)} ~.
\end{equation}
Using the right triangle SCH, we have 
\begin{equation}\label{Eq:cosgamma}
    \cos{\gamma} = \frac{\rs - h_z}{\rs} = 1 - \frac{h_z}{\rs}~,
\end{equation}
where $h_z$ is the length of the segment HZ, or the height of the zenithal cap defined by the position of the satellite at its zenithal distance $z$. Inserting the expression of $\gamma$ from Eq.~\ref{Eq:gamma} into Eq.~\ref{Eq:cosgamma}, we obtain the expression for $h_z$:
\begin{equation}\begin{split}
        h_z &= (1 - \cos{\gamma})\rs ~,\\
        &=  \rs \left( 1 - \cos{ \left( z -\arcsin{\left( \frac{\re}{\rs} \sin{z}\right) } \right) }\right) ~.\\
\end{split}\end{equation}

Now that we have $h_z$, we can compute the area of the spherical cap above $z$,
\begin{equation}
    A_{\rm vis.} = 2 \pi ~ \rs ~ h_z ~,
\end{equation}
and the area of the whole sphere containing the satellite,
\begin{equation}
    A_{\rm total} = 4 \pi ~ \rs^2 ~.
\end{equation}
The number of satellites visible in the cap above $z$ is therefore
\begin{equation}\begin{split}
    N &= N_{\rm Const.} \frac{A_{\rm Vis.}}{A_{\rm Total}}~,\\
    & = N_{\rm Const.} \frac{h_z}{2 \rs} ~,\\
    & = \frac{N_{\rm Const.} }{2} \left( 1 - \cos{ \left( z -\arcsin{\left( \frac{\re}{\rs} \sin{z}\right) } \right) }\right) ~. \\
\end{split}\end{equation}
The value of $N$ is reported for each constellation in Table~\ref{tab:const}, for  $\zmax= 90\dg$ (objects above the horizon) and  $\zmax = 60\dg$. 

\B{Using these formulae, Figure~\ref{Fig:fractions} shows the number of constellations in range above a range of elevations, and the corresponding density of satellites per square degree for the considered constellations. It also displays the fraction of constellations in range as a function of elevation for various satellite altitudes; this plot can be used to estimate numbers for arbitrary constellations. It is interesting to note that while 2--12\% of a constellation is above the horizon, this fraction drops to 1--6\% at 15$\dg$ elevation, and 0.5--3\% at 30$\dg$, for altitudes in the range of 250--2000km.}

\citet{galadie20} performed detailed simulations of the number of satellites in range from various observatories using the actual orbital distribution of the Starlink satellites as Walker constellations, and obtained $N \sim 40$ -- 80 with $z<60\dg$, from low- to high-latitude observatories, respectively. The present geometric approximation finds 54 satellites in the same conditions (independently of latitude). The referee of the present paper also independently obtained similar numbers with a similar method.

\section{Illumination of the satellites}

\begin{figure}
    \centering
    \includegraphics[width=0.4\textwidth]{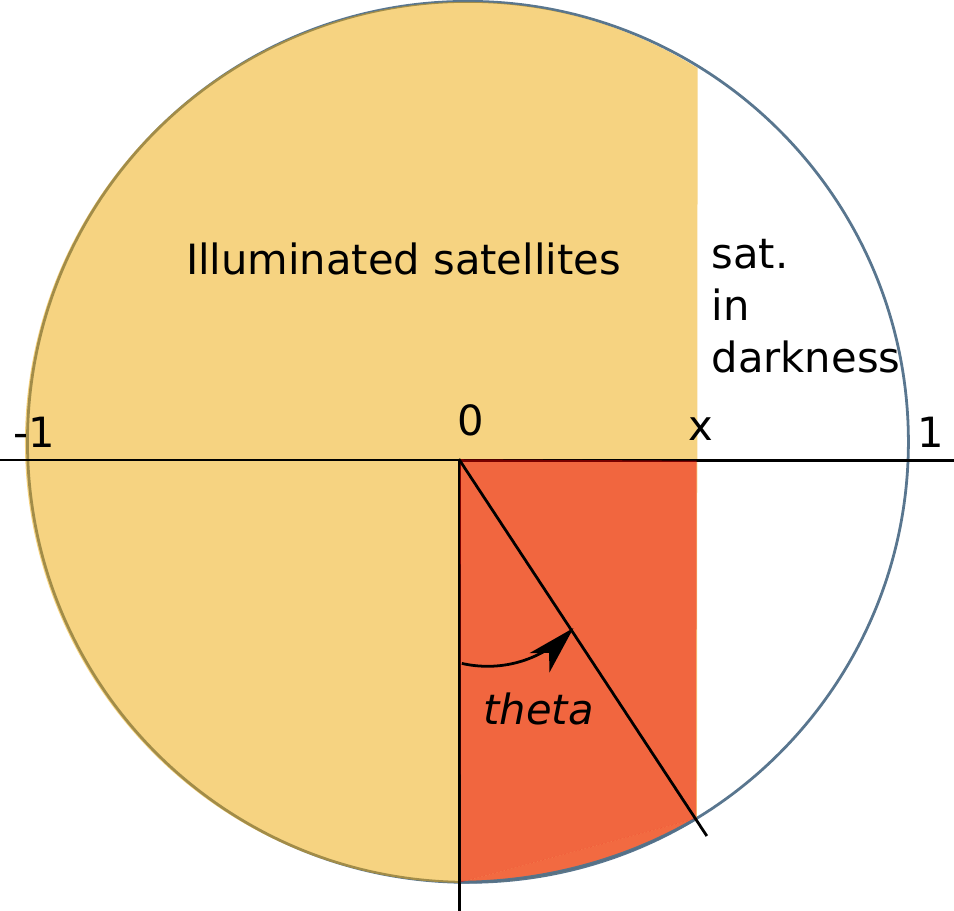}
    \caption{Top view of the sky visible from an observatory; the yellow area indicates the fraction of the sky (for a given orbit altitude) where the satellites are illuminated by the Sun; the area in red is used to compute the area in yellow.}
    \label{Fig:area}
\end{figure}

\begin{figure}
    \centering
    \includegraphics[width=0.45\textwidth]{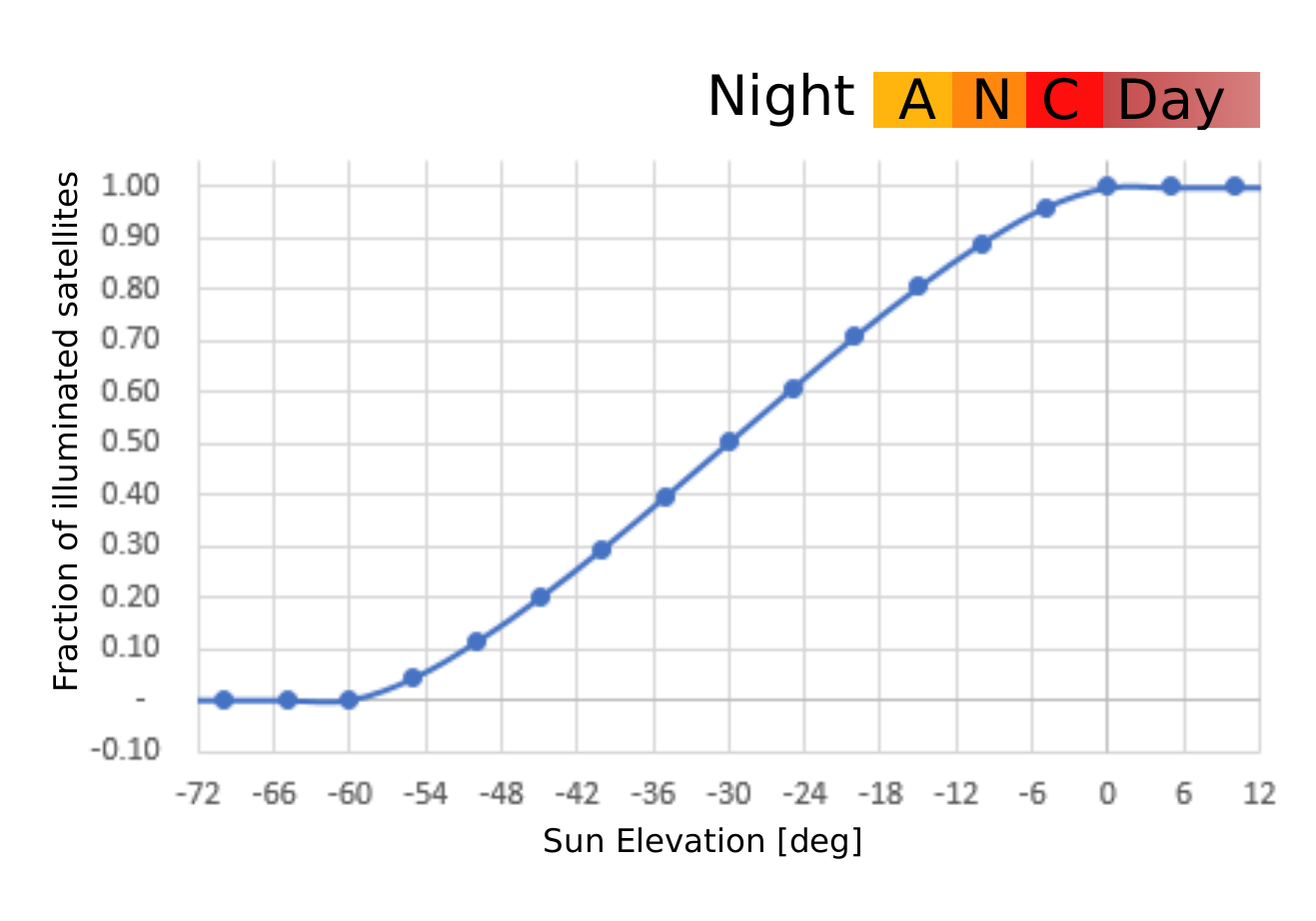}
    \caption{Fraction of the satellites in range and illuminated by the Sun as a function of the Sun's elevation above the horizon. The satellites' orbits have an altitude of 1000~km in this example. The region shaded in dark red corresponds to day time; in red (C) to civil twilight,  when observations are not possible or are not affected by bright sources; in orange (N) to nautical twilight,  when most observations are not possible; in yellow (A) to astronomical twilight, when observations in the IR or short observations in the visible are possible.}
    \label{Fig:SatTwilight}
\end{figure}

\begin{figure}
    \centering
    \includegraphics[width=0.45\textwidth]{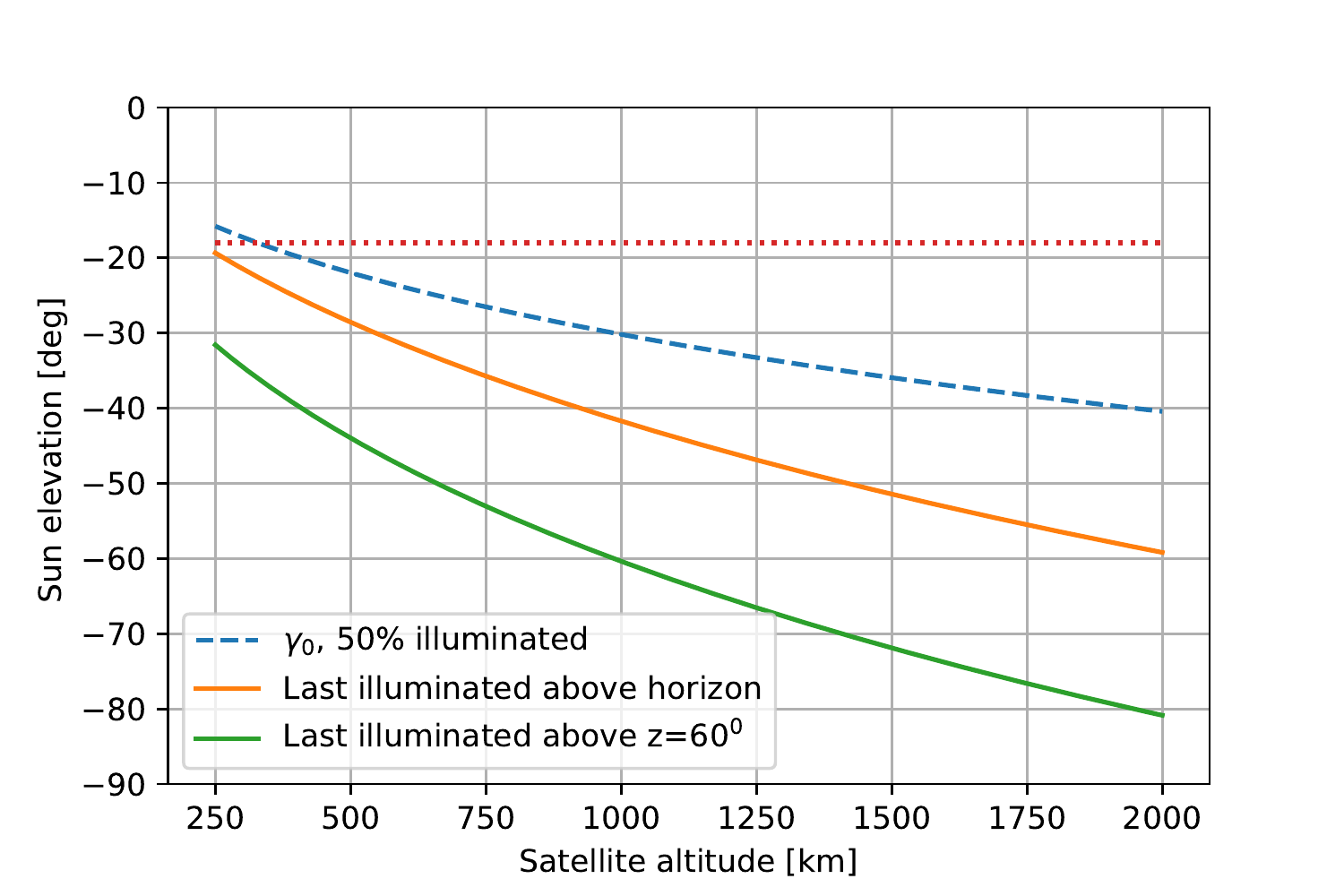}
    \caption{Elevation of the Sun at which half the constellation is in shadow (dashed), and below which the whole constellation is in shadow, considering only those above $z=60\dg$ (orange) or above the horizon (green). The dotted line marks $-18\dg$, i.e. astronomical twilight.}
    \label{Fig:sunElev}
\end{figure}

In the visible and NIR, the satellites are visible only because of reflected sunlight. Therefore, in order to be observable, a satellite must be both in range (above the horizon, or above $z$) and illuminated. 

The fraction of visible satellites that are illuminated by the Sun varies with the Sun elevation below the local horizon. When the Sun is above or on the local horizon, all the satellites above the horizon are illuminated. \B{By construction, the angle $\gamma_o$ introduced above is also the elevation of the Sun below the horizon, just illuminating a satellite at zenith (i.e. when half the satellites in range are illuminated), and $2 \gamma_o$ is the elevation of the Sun below the horizon, just illuminating a satellite on the horizon toward the Sun (i.e. the elevation of the Sun below which no satellite is illuminated). These values of the Sun's altitude are listed in Table~\ref{tab:const}.} 
Generalising, let $a$ be the elevation of the sun below the horizon that just illuminates a satellite at zenithal distance $z$. By construction,
\begin{equation}\label{Eq:a}
    a = \gamma + \gamma_o ~,
\end{equation}
and the expressions for $\gamma(z)$ and $\gamma_o$ are given in Eqs.~\ref{Eq:gamma} and \ref{Eq:gamma_o}.

Simplifying the geometry of the terminator to make it a straight line (acceptable for low earth orbit (LEO) satellites as the spherical cap considered is small compared to the sphere), the fraction is obtained by integrating the fraction of the spherical cap on the Sun-side of the terminator, approximated by 
\begin{equation}
f    = \frac{A_{\rm illuminated}}{A_{\rm total}} ~.\\
\end{equation}
For satellites on higher orbits than those considered here, and for observatories at high latitudes and observations far from the equinox, the actual shape of the shadow cone of the Earth will not satisfy this approximation, but it is acceptable for LEOs. From Fig.~\ref{Fig:area}, $ A_{\rm illuminated} = \pi/2 + 2 A_{\rm red}$ where $ A_{\rm red}$ is the area highlighted in red. With $\theta$ the angle from the centre so that $X = \sin \theta$, we have
\begin{equation}
\begin{split}
    A_{\rm red} &= \int_0^X \cos( \arcsin(x)) dx\\
    & = \frac{1}{2} \left[ x \sqrt{1-x^2} + \arcsin x \right]_0^X\\
    & = \frac{1}{2} \left( X \sqrt{1-X^2} + \arcsin X \right) ~.
\end{split}
\end{equation}
Using this in the equation for $ A_{\rm illuminated} $, and with $A_{\rm total} = \pi$, we have
\begin{equation}\label{Eq:fraction}
f =  \frac{1}{2} - \frac{X \sqrt{1-X^2} + \arcsin{X}}{\pi} ~. \\
\end{equation}

\begin{figure*}
    \centering
    \includegraphics[width=0.85\textwidth]{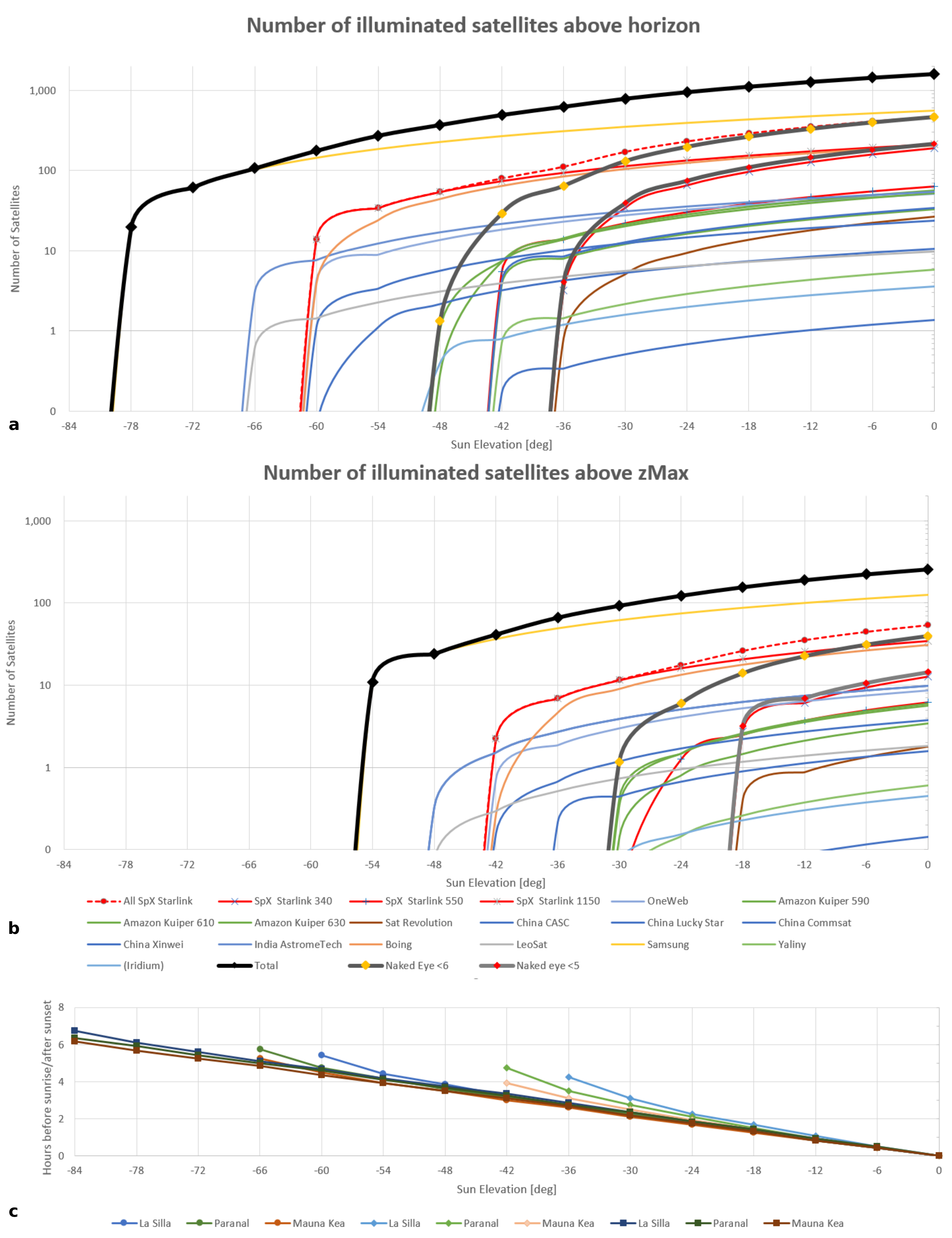}
    \caption{Number of satellites in range (\textbf{(a)} above the horizon, or \textbf{(b)} above zenithal distance $z=60\dg$ or elevation $e>30\dg$)  and illuminated by the Sun, as a function of the Sun's elevation. The individual constellations are represented, as well as the whole of the SpaceX constellations (dotted red line), and the grand total of all considered constellations (thick black line). The number of satellites visible to the naked eye is also indicated (\B{mag brighter than 6, thick grey line with yellow symbols, or mag brighter than 5, thick grey line with red symbols}).  The number of satellites visible with the naked eye is around 100 at the beginning and end of the astronomical night, an order of magnitude less are visible above $30\dg$ elevation, and this number plummets when the Sun drops below $50\dg$.
    Only some observations are possible for sun elevations $>-18\dg$ (astronomical twilight), even fewer observation types (and shorter ones) are possible for $>-12\dg$ (nautical twilight), and virtually none for $>-6\dg$ (civil twilight).
    The bottom plot (c) indicates the number of hours before and after sunrise and sunset corresponding to the sun altitude, for three observatories, for both solstices and an equinox.
    }
    \label{fig:SatNumber}
\end{figure*}

The function in Eq.~\ref{Eq:fraction} is displayed at Fig.~\ref{Fig:SatTwilight}. It is approximated by a linear function as
\begin{equation}
\begin{split}
            f_{\rm Illuminated} & = 100\%  {\rm ~~~if~~~} a_\sun \ge 0\\
            & = 1 - \frac{a_\sun}{a(\zmax)} {\rm ~~~if~~~} 0 \ge a_\sun \ge a(\zmax) \\
            & = 0 {\rm ~~~if~~~} a_\sun < a(\zmax)  ~, \\
\end{split}
\end{equation}
where $a_\sun$ is the actual elevation of the Sun. \B{Figure~\ref{Fig:sunElev} shows, as a function of the altitude of the satellites, $a_\sun = \gamma_0$ corresponding to 50\% of the illuminated satellites, and $a_\sun$ corresponding to the whole constellation in range being in the Earth's shadow, considering the case of $z=60\dg$ and $z=90\dg$ (horizon).}
These fractions are applied to all the constellations for various sun elevations from sunset until beyond the time when all satellites are in  shadow. The results are displayed in Fig.~\ref{fig:SatNumber} for each constellation and for the complete collection.

The correspondence between the elevation of the sun and time is represented in Fig.~\ref{fig:SatNumber}.c. Thanks to the uniform satellite distribution, these fractions are valid for any observatory, and for any date. The linear approximation for the terminator restricts their validity to satellites on orbits below a few thousand kilometres, which is valid for the constellations considered. \citet{galadie20} computed the fraction of illuminated Starlink satellites using actual Walker constellations; our results are in agreement with his. The seasonal effects determined by this latter author for low- to mid-latitude observatories are small, confirming the validity of our simplifying approximations. 

It is worth noting that satellites on very low orbits are illuminated during only a brief period immediately after sunset and before sunrise. Because of this, the trains of satellites on their very low transfer orbit immediately after launch are visible only briefly during twilight.

\section{Magnitude and brightness of the satellites}

The satellites are complex objects, with a complicated reflection and diffusion behaviour. Flat, polished panels (such as some of the satellite's body or the solar panels) act as mirrors, causing specular reflections that, when pointing toward Earth, affect a very small area of the planet but can cause an extremely bright flash. Other parts of the satellites will diffuse light. Furthermore, the satellite attitude with respect to the Sun and the observer will complicate matters. In the spirit of this paper, this is simplified using a straightforward model of this complex situation.

\subsection{Visible magnitude and diffusion}

The satellite is simplistically represented by a sphere, characterised by its radius $r$ and its albedo $p$. We consider only simple geometric diffusion (i.e the diffused light is proportional to the cross-section of the object), and the solar phase effect for the sphere is described by the phase function:
\begin{equation}
    f(\alpha) = \frac{1+\cos \alpha}{2} ~,
\end{equation}
where $\alpha$ is the solar phase angle. The magnitude of the object is then
\begin{equation}\label{Eq:magnitude}
    M = M_{\rm Sun} - 2.5 \log( f(\alpha) r^2 p ) + 5 \log( R \Delta ) + x \chi ~,
\end{equation}
where $M_{\rm Sun} = -26.75$ is the magnitude of the Sun (in the $V$ band, around 550nm);
$r$ is the radius of the object expressed in astronomical units (1 au $= 1.495979 10^{11}$~m);
$R$ is the heliocentric distance of the object, 1 au;
$\Delta$ is the distance between the object and the observatory, also expressed in astronomical units. 
The term $x \chi$ represents the absorption by the atmosphere, where $x\simeq 0.12$~mag/airmass is the extinction (0.12 is a typical value in the visible filter $V$; see e.g. \citet{patat+11}) and $\chi = 1/\cos(z)$ is the airmass, which is the quantity of atmosphere crossed by the observed light, normalized to zenith. This equation is customary for the magnitude of asteroids; see for example \citet{gehrels+79}. Here, $\Delta$ is obtained
from the zenithal distance of the satellite, 
\begin{equation}\label{Eq:zd}
    z = \arctan \left(  \frac{\re - \rs}{\re + \rs} \cot(\gamma/2) \right) + \pi/2  + \gamma/2 ~,
\end{equation}
in
\begin{equation}
    \Delta = \rs \frac{\sin \gamma}{\sin z}~.
\end{equation}

\begin{figure}
    \centering
    \includegraphics[width=0.45\textwidth]{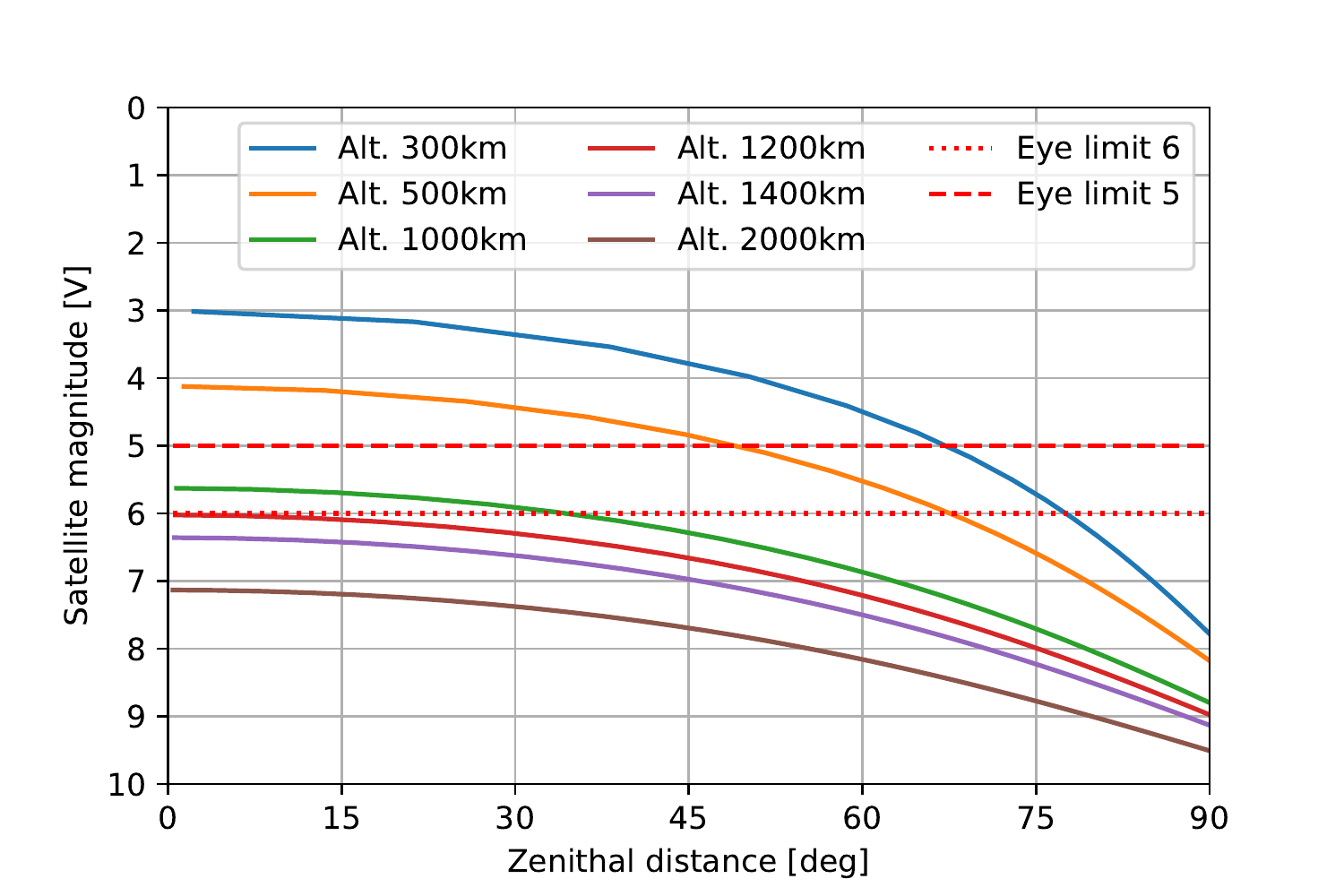}
    \caption{Visible magnitude of Space-X-like satellites as a function of their zenithal distance, for various orbit altitudes. Only objects with a magnitude above the dashed line can be seen with the unaided eye in good conditions (mag 5) and above the dotted line in exceptional conditions (mag 6).}
    \label{fig:satMag}
\end{figure}

The radius $r$ and albedo $p$ of the satellite are difficult to estimate. Measurements of NOAA satellites (1500kg, 3.7$\times 1.88$, mag 4.1 at zenith) indicate that $r=1.5$m and $p=0.25$ reproduce the brightness of the satellite well. Scaling down to the Starlink satellite (550kg) we use $r=1$m and $p=0.25$. This results in a range of  4.2 to 5.9 mag  for Starlink 550km, which is in agreement with a direct photometric measurement of $V=5$ for such a satellite (T.Tyson, priv. comm.). With this assumption, only  objects at the lowest altitudes are visible to the naked eye. The corresponding magnitudes are displayed in Fig.~\ref{fig:satMag}. \B{More recent measurements of the Starlink satellites on their final altitude and attitude indicate they could be as faint as $\sim$ 8 mag; furthermore, Starlink is experimenting with a darkened coating that could make the satellites even darker. We keep the above-mentioned estimate as a conservative, brighter limit, also accounting for the fact that other satellites could be brighter than those of Starlink.}

Using these values, the post-launch low-altitude SpaceX Starlink satellites would appear between  $-2$ and $-1$ mag, in good agreement with the numerous spotting of the Starlink trains. These bright magnitudes combined with the spectacular `string of pearls' appearance of these trains explain the attention they have received.

With these assumptions, and considering that  all the satellites have the same characteristics, the magnitudes of the satellites are listed in Table~\ref{tab:const} for an observation at zenith and at $z=60\dg$. The total number of objects in range, illuminated by the Sun, and visible with the naked eye (mag $<5$ and $<6$) is displayed as a function of the elevation of the Sun in Fig.~\ref{fig:SatNumber}(a and b).

\subsection{Specular flares}

\begin{figure}
    \centering
    \includegraphics[width=0.45\textwidth]{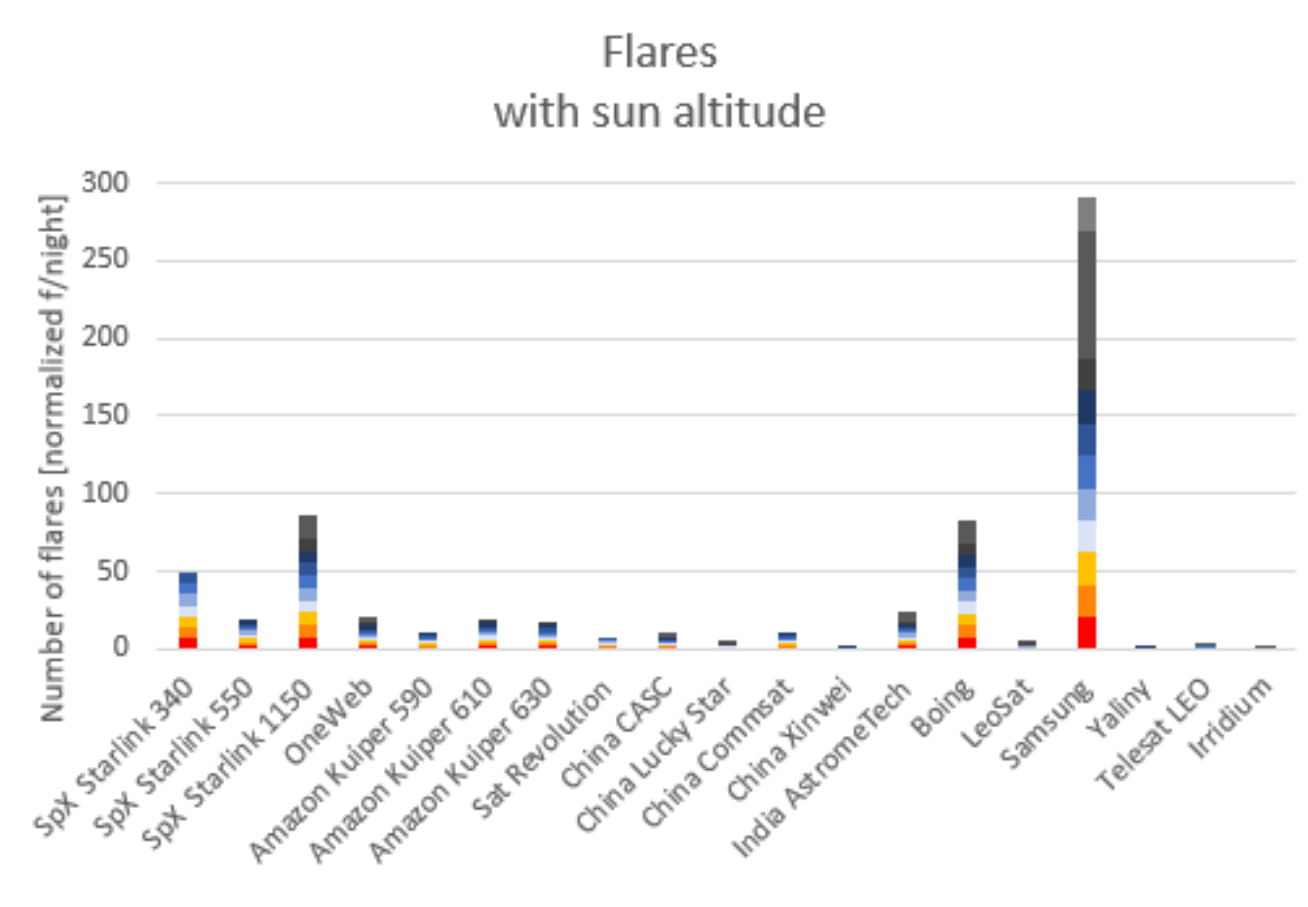}
    \caption{Number of flares for each constellation, simply scaling them to one-third of the flares caused by the original Iridium satellites (which had three large antennas) and to the number of satellites. This is the number of observable flares per night, or the number of flares per week brighter than  $-5 $ mag  for a mid-latitude site. The colour encodes the sun elevation below the horizon, from $0\dg$ (red), $-18\dg$ (pale blue), and into the night (darker blue to greys).}
    \label{fig:flares}
\end{figure}

The original Iridium constellation was well known for its spectacularly bright flares, where each of their three $\sim 1\times2$m antennas illuminated a $\sim$10~km diameter on the ground. With the 66 satellites on 800~km altitude orbits, Iridium flares were visible quite often (two to four times per night). Flares of  -5 mag in brightness occurred three to four times per week; flares of -8 mag may be visible three to five times per month for stationary observers (Wikipedia). Newer Iridium satellites do not cause noticeable flares. It is unknown which, if any, of the new satellites in the LEO constellations will cause flares. As a conservatively pessimistic approach, we consider that every satellite will have one Iridium-like reflecting surface that causes flares similar in brightness and frequency to those caused by Iridium's antennas. This is extremely pessimistically conservative. Simply scaling the flare frequencies to one-third of those caused by Iridium (where each satellite had three antennas) and to the number of satellites leads to a total of about 660 flares visible
above the horizon per night including 100 flares brighter than  $-5$ mag. Above $z=60\dg$, these numbers convert to 100 flares per night, including 20 brighter than mag $-5$. Assuming that the flares occur at random times while the satellite is illuminated, Fig.~\ref{fig:flares} displays the contribution of each constellation for different solar elevations.

\subsection{Thermal infrared emission}

In the 5--20$\mu$m range, the satellites will emit a considerable amount of thermal IR radiation. Simulating the details of that emission would not be simple: the surface materials are chosen to maintain the temperature of the satellite within the operational range, and include thermal radiators designed to eliminate the heat generated by the instrumentation and received from the Sun and from the Earth (the Earth-facing side receives significant thermal radiation from Earth due to the large viewing angle). 

\B{Simplifying this to the extreme, a satellite is represented by a sphere with a   diameter of 1m and an albedo of 0.25. An emissivity of 0.1 leads to temperatures over 400K, probably not realistic for the hardware. Using an emissivity of 0.9 leads to an isothermal temperature of $\sim 300$~K; for a satellite at an altitude of 2000~km, this would produce a flux of up to 100~Jy in N-band (8--13~$\mu$m), and several tens of Janskys in the M- and Q-bands (5 and 18-20~$\mu$m, respectively; Th. Mueller, priv. comm.).}

As the satellite operators are certainly striving to keep the inside of the satellite at $\sim 300$~K, and as the satellites alternate between solar illumination and Earth shadow on an hourly basis, we consider that the temperature of the satellite is constant, and that the thermal IR flux is constant at 100~Jy in N-band and 50~Jy in M- and Q-bands. \B{These are conservative estimates --the actual flux could be significantly lower}.
The number of satellites relevant for thermal IR observations is then simply the number of satellites in range; whether or not they are illuminated by the Sun is irrelevant.


\section{Observation contamination}\label{sub:trails}

\begin{figure}
    \centering
    \includegraphics[width=0.45\textwidth]{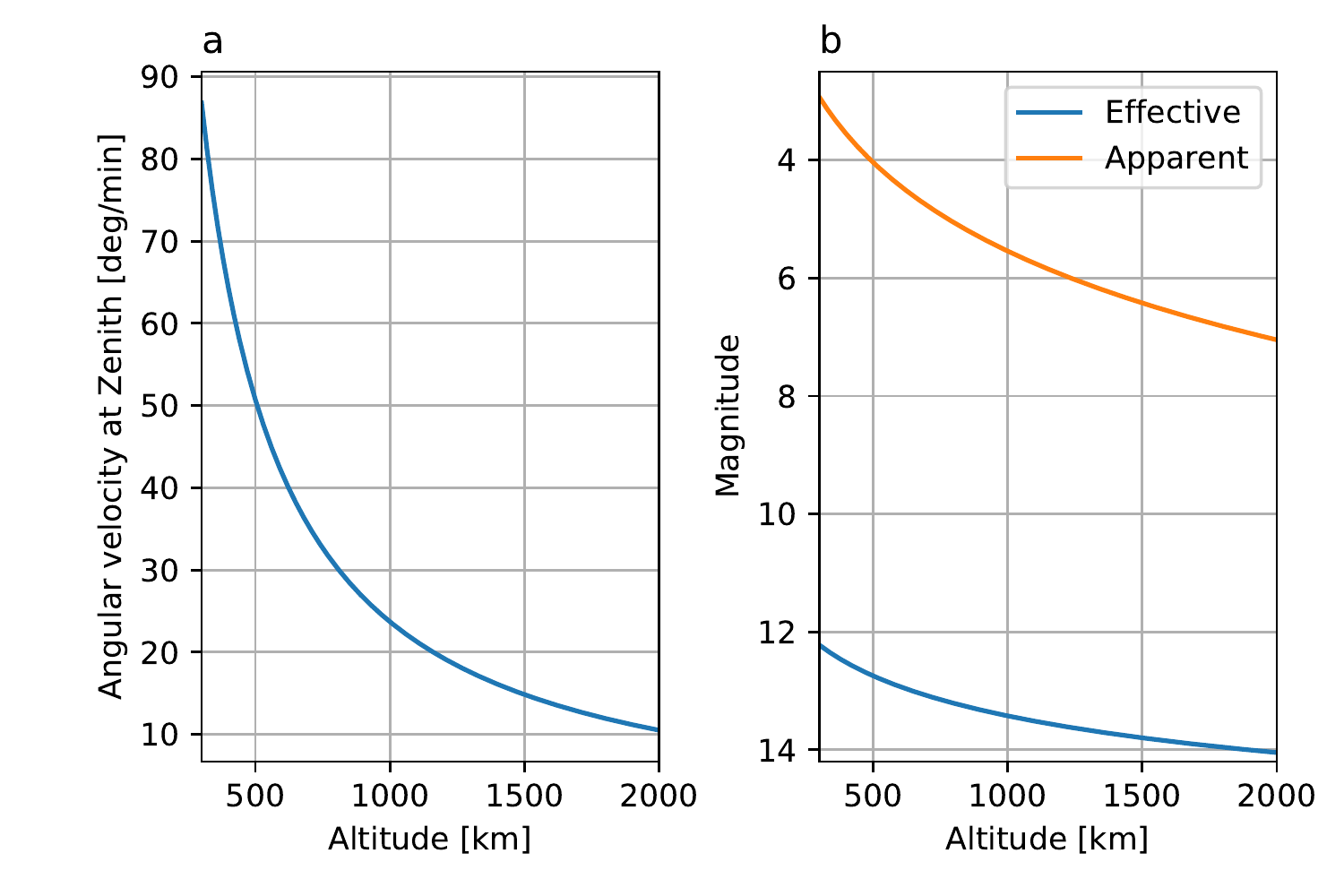}
    \includegraphics[width=0.45\textwidth]{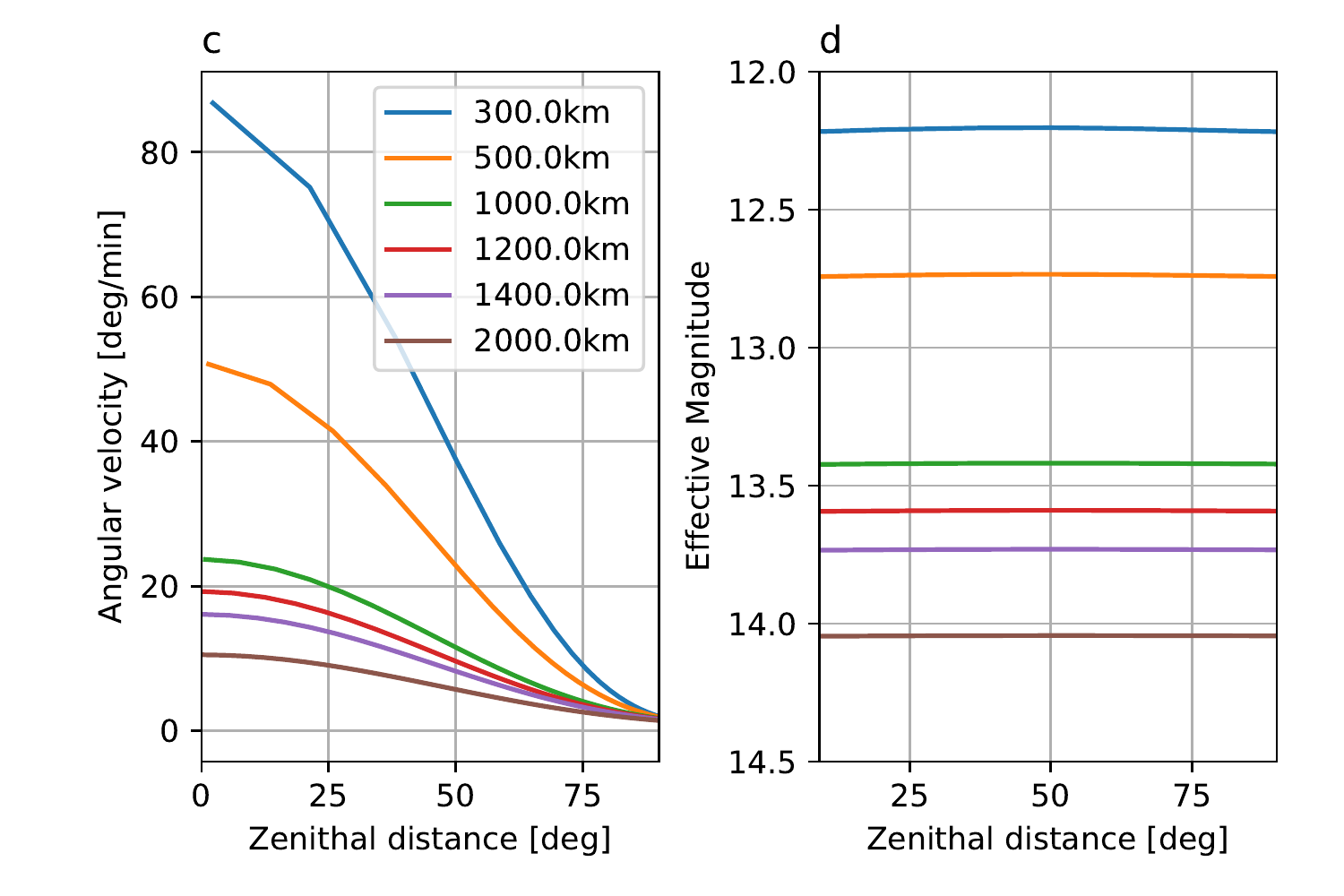}
    \caption{\B{(a) Angular velocity of a satellite as a function of its altitude, around Zenith; 
    (b) Again at Zenith, effect of the altitude on the apparent magnitude of the satellite, and on the effective magnitude accounting for trailing (with a seeing of 1 arcsec);
    (c) Angular velocity of a satellite as a function of its zenithal distance, for various altitudes;
    (d) Combining the effect of distance and trailing on the effective magnitude, showing no dependency with zenithal distance.}
    }
    \label{fig:satVel}
\end{figure}

\begin{figure}
    \centering
    \includegraphics[width=0.45\textwidth]{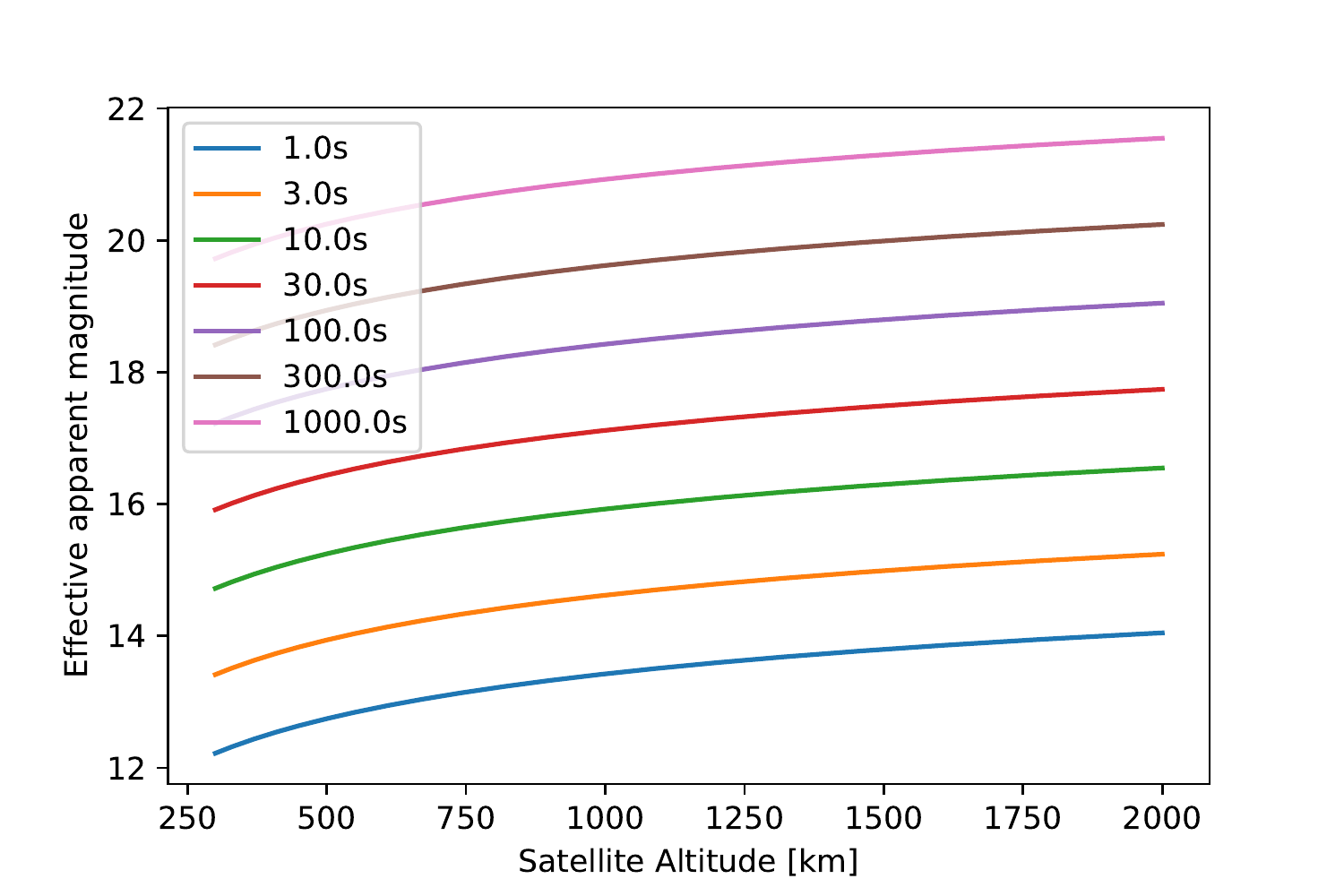}
    \caption{Effective apparent magnitude of the satellite as a function of its altitude, for various exposure times. A field star with that magnitude will have the same peak brightness as the satellite in that exposure. }
    \label{fig:effAppMag}
\end{figure}

One way to evaluate the fraction of observations affected by satellites would be to compute the position of all the satellites in the sky above an observatory at a given time, `shoot' a series of exposures, and compute how many of these have a satellite in the field of view. We instead used a geometrical probabilistic approach: considering the duration of an observation (the individual exposure time), we estimate the fraction of the sky that is covered by satellite trails during that exposure time. The field of view of the observation is accounted for by setting the width of the satellite trails. In that way, we immediately have an estimate of the probability of having an exposure affected by a satellite: this is estimated as the fraction of the sky covered by satellite trails.
We consider various types of scientific exposure times over a representative set of ESO instruments.
\begin{itemize}
    \item Standard imaging in the visible (e.g. with FORS2 or EFOSC2), or the NIR (e.g. with HAWKI). Individual exposure times range from a few seconds to a couple of minutes for broad-band filters, and to several minutes in narrow-band filters. For the simulation, we use an exposure time of 100s, and a field of view of 6$'$ in diameter.
    \item Wide-field imaging in the visible (e.g. with OmegaCam) or in the NIR (e.g. with VIRCAM). Exposure times are similar as in the previous case. We use 100s, and a field of view of 1$\dg$. 
    \item Long-slit spectroscopy, in the visible (e.g. with FORS2) or the NIR. Typical exposures range from a few minutes to one hour. We use 1000s, and a slit length of 6$'$. 
    \item Short-slit spectroscopy, in the visible (e.g. with UVES or XSHOOTER) or the NIR (e.g. with CRIRES+ or XSHOOTER). Typical exposures range from a few minutes to one hour. We use 1000s, and a slit length of 12$"$. 
    \item Fibre-fed spectroscopy in the visible (e.g. with HARPS or ESPRESSO) or the NIR (e.g. with NIRPS). Typical exposures range from a few minutes to one hour. We use 1000s, and a fibre diameter of 2$"$.
    \item Multi-fibre spectroscopy, for example with FLAMES or 4MOST. Typical exposures range from a few minutes to one hour. The fibres are positioned over a broad but very sparsely populated field of view. 4MOST has 2400 fibres on a 4.6 sq.deg field of view with a 2.6$\dg$ diameter. 
    In the worst case, up to 30 fibres could be affected by a satellite trail. The effect on this instrument is obtained by multiplying the effect on one fibre by 30.
\end{itemize}

For spectrographs, we consider that the slit is always perpendicular to the motion of the satellite, conservatively maximising the cross-section. 

To estimate the length of the trail left by a satellite during an exposure, the observed angular velocity is obtained by computing numerically the derivative of the zenithal distance (from Eq.~\ref{Eq:zd}) accounting for the orbital velocity of the satellite using Kepler's law.

\B{The apparent angular velocity is a function of the altitude (an object being further away appears to move more slowly because of the slower intrinsic motion and the larger distance; see  Fig.~\ref{fig:satVel}(a)), and of the zenithal distance (the lower the object, the slower its apparent motion because of foreshortening; Fig.~\ref{fig:satVel}(c)).  The effective magnitude of a satellite will depend on the distance between the satellite and the observer, which is a function of the zenithal distance of the satellite, and on the trailing of the satellite during the exposure, which is a function of its angular velocity. For a typical seeing of 1~arcsec, the length of the trail in arcsec will give the attenuation factor in magnitude, $2.5 \log( v )$ ($v$ in arcsec/sec). The geometric effect and trailing attenuation effect are illustrated in Fig.~\ref{fig:satVel}(b). The additional effects of the zenithal distance on the geometric distance and on the apparent velocity compensate each-other, as seen in Fig.~\ref{fig:satVel}(d).   
As a reference, Fig.~\ref{fig:effAppMag} displays the  apparent effective magnitude of the satellite as a function of its altitude, for various exposure durations. 
}

The width of the contamination trail depends on the magnitude of the satellite and the observing technique:
\B{\begin{itemize}
    \item Bright flare (mag $< 0)$: The satellite trail heavily saturates the detector. We consider that the whole field of view of the instrument is contaminated by the trail, either directly, by spurious reflections and diffusion of the light from the trail in the instrument, or possibly by contamination caused by cross-talk or interference in the saturated electronics of the instrument. 
    \item Medium-brightness satellites ($0<$mag$<5$) leaving a saturated trail on the detector. Conservatively, we assume the whole field of view is contaminated.  
    \item Fainter satellite (mag $>5$) leaving a non-saturated trail on the detector: the track width will extend over a few times the seeing, conservatively set to $5"$ for imagers and long-slit spectroscopy, and to the full slit in case of a short slit. 
\end{itemize}
}

The boundary values between the brightness categories are a function of the diameter of the telescope and of the sensitivity, dispersion, and transmission of the optical elements: a low-efficiency spectrograph on a small telescope will indeed be less affected than an imaging camera with broad-band filters on a giant telescope. The values chosen are representative of a large telescope like ESO's 8m Very Large Telescope (VLT), but are also valid for smaller telescopes like the ESO 3.6m New Technology Telescope and the upcoming 39m Extremely Large Telescope (ELT). 

Because of the extreme case of the Rubin Observatory (formerly known as LSST, with a large diameter of 8m, high sensitivity of the detectors, and gigantic field of view of 10~sq.degrees), we consider it separately.  Based on reports by Tyson (priv.comm.), the effect of a bright satellite trail contaminates the full field of view, and that of a fainter satellite contaminates a full quadrant of the instrument (either directly, or through electronic cross-talk in the camera electronics causing unremovable signal). LSST observes typically with an exposure time of 30s. For the simulations presented below, the field diameter is set to 3.5$\dg$ for all satellites brightnesses, which constitutes a pessimistic limit: a faint satellite would potentially affect only a quadrant of the camera.
The estimates presented below scale with the exposure time and the field of view, and therefore the effects can be adjusted to other instruments and exposure times.

The area $A$ of the sky covered by satellite trails is obtained from
$A = t v w N$, where $t$ is the exposure duration in seconds, $v$ the angular velocity of the considered satellites in deg/second, $w$ the width of the field of view for the considered observation type in degrees, and $N$ the number of considered satellites in range and illuminated. The contributions of the various types of satellites are summed, resulting in the total contaminated area. As $N$ is a function of the elevation of the Sun below the horizon, the computation is repeated for various bins of solar elevation. If the length of a trail $t v$ is too long to fit in the observable sky, this simply means that the first satellite disappeared over the horizon, and was replaced by a new one entering the observable sky. The total area of the observable sky above $z=60\dg$ is $A_{\rm sky} = (1 - \cos(60\dg)/2 \times 41\,252.96$~sq.deg = 13\,323~sq.deg. Overlapping satellite trails are counted separately, resulting in an overestimation of the contaminated area (ultimately, this could result in an estimated contaminated fraction $>100$\%). 

In the case of flares, it is assumed that the duration of the flare is $t =10$~s. The number of (10s) flares at a given time is computed scaling the frequencies (in number of flare by night) to the duration of a flare, accounting for one night, which is equal to 10h or 3600 times the duration of a flare.

In the case of thermal IR emission, the effect of the satellites does not depend on them being illuminated or not, so the contaminated fraction of the sky does not change with solar elevation. 

The contaminated fraction directly gives the probability that a given exposure will be lost due to a satellite; these  are listed in Table~\ref{tab:contamination}. We note that these fractions scale linearly with the exposure times and the field of view, meaning that the effect on other specific exposures can be inferred from this table. 

\subsection{Contamination of observations in the visible and near-infrared by specular flares}

Even considering very pessimistic estimates (i.e. each satellite has one Irridium-like reflecting surface) and considering the complete collection of constellations with their conservatively pessimistic uniform distribution, only long exposures (1800~s) with wide-field survey cameras (1 sq.deg) would be marginally affected (at the $10^{-4}$ level). One must note that this type of exposure is extremely rare (images have rarely an exposure time longer than 10~min). All the other categories of exposures are affected much below the $10^{-4}$ level. Specular flares are therefore not considered an issue for telescopic astronomical observations.

\subsection{Contamination of observations in the visible and near-infrared by satellite trails}
\B{
The effect of satellite trails is different for the various types of exposures considered:
\begin{itemize}
\item \textbf{Short exposures} (1s) are essentially not affected by the satellite trails. 
\item \textbf{Medium-duration exposures} (100s) are affected at a very low level (below 0.1\%) during the night, and at a low level (0.5\%) during nautical twilight.
\item \textbf{Long spectroscopic exposures} (1000s) are affected at less than the \% level during the first and last couple of hours of the night, and at the 1\% level during astronomical twilight. This can --in most cases-- be mitigated by not scheduling long exposures during the astronomical twilights (which are usually not suitable for these observations anyway) and the first and last hour of the night.
\item \textbf{Wide-field imaging (OMEGACAM) and multi-fibre spectrographs (4MOST)} are affected at the 5--7\% level at the beginning and end of night.
\item \textbf{LSST} ultra-wide exposures on a large telescope: up to 30\% of the exposures would be lost during the first and last hours of the night, and almost 50\% of the twilight exposures would be contaminated. The combination of wide field of view and the huge collecting area of a large mirror makes this type of observation very sensitive to satellites. This is likely to cause significant disruption in the scheduling and efficiency of the surveys.
\item \textbf{Caveat}: Except in the case of the LSST, these estimates consider that a long-slit spectroscopic frame or an image affected by a faint satellite (mag fainter than 5, effective magnitude below 16) is not completely ruined by the trail, and that the remaining part of the frame can be used for science, for instance by combining it with other frames. This will not be true for all science cases: for some programmes, any trail in the field of view could ruin the whole frame, no matter how faint it is. For these science cases, the fraction of affected exposures could be in the 10--20\% level around twilight depending on the exposure time and field of view, and mitigation measures would be needed.
\end{itemize}
}

\begin{table*}[]
    \centering
    \includegraphics[width=\textwidth]{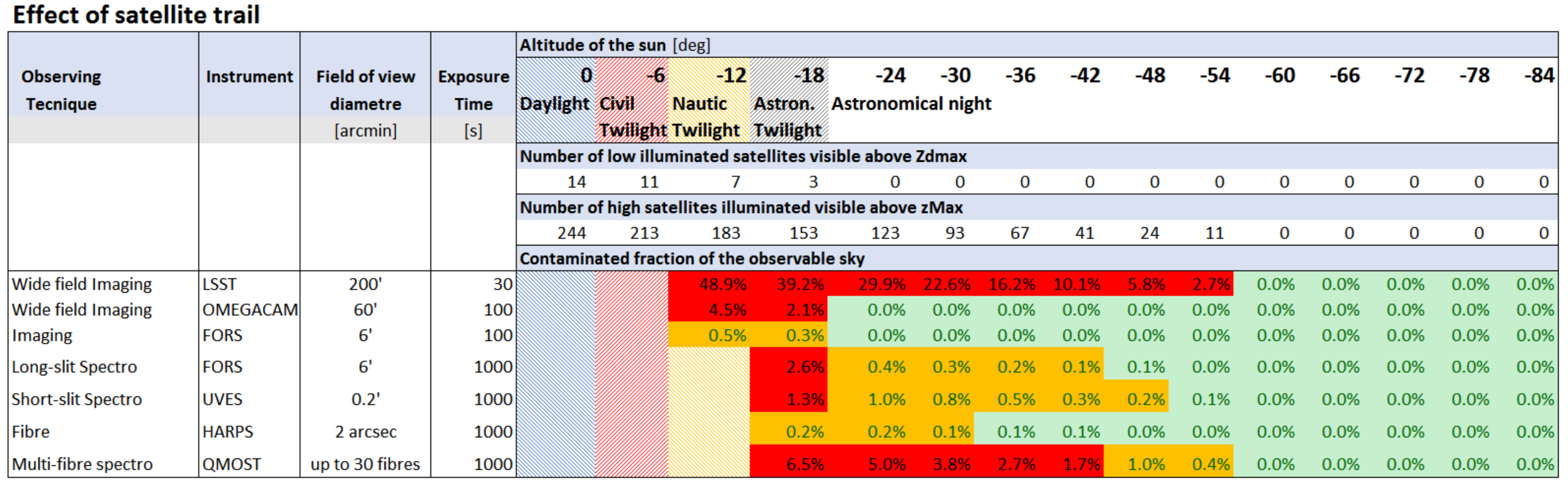}
    \caption{Probability that an exposure is ruined by a satellite trail, expressed as the fraction of the observable sky (down to $z=60\dg$) contaminated by at least one trail during the duration of the exposure, for the considered observing technique. These are listed as a function of the elevation of the sun (in degrees below the horizon). Low- and high-altitude, and bright and faint satellites are evaluated separately and combined in these totals.  
    Various observing techniques are considered, each having a different field of view and typical exposure time. In the case of LSST, because of the heavy saturation of the satellite trails, it is assumed that the whole field of view is entirely ruined by a satellite.
    }
    \label{tab:contamination}
\end{table*}


\subsection{Contamination of thermal infrared observations}

The signal in ground-based thermal IR observations is dominated by the thermal emission of the sky and of the telescope, the astronomical component being a small addition to that bright background. This requires extremely short individual exposures (a fraction of a second, e.g. 0.02~s for VISIR on the ESO VLT), and an observation method using chopping (typically by moving the secondary mirror of the telescope) at a few Hertz, and nodding the whole telescope every few seconds. During one of these 0.02~s individual exposures, a static 100~Jy source would be detected with a signal-to-noise ratio (S/N) of $\sim 100$ with VISIR. However, the image of the satellite on the detector will be trailed by an amount that depends on its altitude, and the flux will be scaled with the inverse square of the altitude. Overall, a satellite would in all cases leave a highly visible trail, with an S/N of 9, 12, and 50 for a satellite at 2000, 1200, and 300~km, respectively. However, these trails are sufficiently faint  compared to the bright background that they would not have additional side effects.

Using again VISIR as an example, with a field of view of $38 \times 38"$, the very short individual exposure time results in extremely low probabilities that an individual exposure will be affected by a satellite trail (about $10^{-6}$ during civil twilight). However, in most observing modes the individual exposures are not saved separately but are combined, averaging all the data acquired on one nodding position, resulting in $\sim 10$s. The probability that at least one of the individual exposures composing that average is contaminated is of the order of 0.1\% during civil twilight. Overall, thermal IR observations are therefore not significantly affected by the emission of the satellites.


\subsection{Occultations}
When a non-illuminated satellite passes in front of an astronomical source, it will briefly occult the light.
The angular size of the satellite, arcsin( diameter/distance), is of the order of 0.2--1$"$.
The apparent angular velocity of the satellites (see Fig.~\ref{fig:satVel}) is of the order of 15 to 80~deg/min.

The occultation duration for a point-source is therefore in the $2\times 10^{-4}$ to $1\times 10^{-3}$~s range.
Using the same mechanism as in Sect.~\ref{sub:trails}, the field of view of an occultation is 1~sq.arcsec (i.e. a conservative value for the angular size of the satellite), and the width of the trail is set to 1$"$. The exposure times considered are 10, 1, and 0.1~s. Accounting for low- and high-orbit satellites, the probability of one exposure being affected by an eclipse is $\sim 10^{-4}$, $0^{-5}$, and $10^{-6}$ (respectively). The effect of the eclipse ranges from $2\times10^{-5}$~mag (10s exp, low satellite) to $1\times10^{-2}$~mag (0.1~s, high satellite). 

Overall, the effect therefore ranges from negligible to small (10 mmag is about the limit of what can be measured from the ground). The probability of these occultations occurring is small: at worst, $\sim 10^{-4}$ of 10~s exposures affected, or about one 10~s exposure every three nights of observation.


%


%

\section{Mitigation measures}
Two main types of mitigation can be considered. The first is scheduling of the observations: At the global level, observing toward the direction opposite to the Sun (toward the east in the evening and toward the west in the morning) will ensure that the satellites are in the shadow of the Earth, therefore avoiding contamination of the exposures. While this is simple to implement and will work even for a wide field of view, this mitigation method is not suitable for all programmes. At a much more detailed level, it is possible to forecast the position of the satellites from their orbital elements, and to observe a field at a time when it will not be crossed by a satellite. The implementation of this mitigation is much more complex, and is not suitable for all programmes (e.g. it may turn out to be impossible to schedule a long exposure with a wide field of view). 

The second type of mitigation involves interruption of the observations: For programmes that require observations in the region of the sky where the satellites are illuminated, it is possible to compute the exact time when a satellite will cross the field of view, and close the shutter during that time. The implementation of that mitigation would be complex, and is not suitable for all programmes (e.g. a large field of view could require so many interruptions that the exposure would not be practical). In both cases, the availability of high-precision, up-to-date orbital elements for all the satellites would be crucial so that the accurate position and timing of the satellites can be computed.

\section{Summary}

This study presents a very simple evaluation of the effect of mega-constellations of low-altitude satellites on telescopic astronomical observations in the visible and IR wavelength domains. The main simplifications are $(i)$ a uniform distribution of the satellites over the globe, $(ii)$ a simple --but empirically calibrated-- model for the brightness of the objects, and ($iii)$ a geometric probabilistic approach of the contamination.
Because of the very drastic simplifications of the problem, its results have to be considered as order-of-magnitude estimates, and will need to be refined using detailed simulations including the actual satellite orbits, a refined photometric model of the satellites (ideally tuned for the various satellite models across constellations), a less crude description of the effect of a satellite trail on the data, and so on. Nevertheless, as most approximations are conservative, and as the number of satellites considered is very large, the presented results are likely to err on the pessimistic side.

This study considers only the visible and IR regimes. A separate paper will deal with the millimetre and submillimetre domains. The radio domain is also to be considered separately.
Keeping in mind the limitations of this study, one can already draw the following conclusions for when the 26\,000 satellites from 18 representative constellations are launched and are in operation:
\begin{itemize}
    \item About 1600 satellites will be in range (over the horizon) of an observatory at mid-latitude. Among those about 250 will be above an elevation of  $30\dg$ above the horizon (i.e. in the part of the sky where observations take place). At the end of the evening, that is, in astronomical twilight, or at the beginning of the morning, astronomical twilight (i.e. when the sky is dark for deep astronomical observations), the number of illuminated satellites will be around 1100 above the horizon, and 150 above $30\dg$ of elevation. Of these, about 260 satellites will be bright enough to be visible with the naked eye in exceptional conditions (mag 6 or brighter); about 110 in good conditions (mag 5 or brighter). Most of them will be near the horizon, with up to about 10 above $30\dg$ of elevation --contrary to claims published online that ``satellites will outnumber the visible stars''. These numbers plummet as the Sun drops further below the horizon.
    \item The trains of satellites, forming a bright `string of pearls', brightly visible right after launch, are not an issue for telescopic observations: while they are spectacular, they are very short-lived and visible only briefly after sunset or before sunrise. 
    \item Specular flares, while potentially spectacular (Iridium's ones could reach mag -8), are rare and short enough so that their effect on telescopic observations will be negligible even accounting very pessimistically for one reflecting surface per satellite. The occultation of an astronomical source by a passing satellite has a very low probability of occurrence, and the effect is below the precision of the measurement.
    \item Short telescopic observations (with an exposure time of $\sim 1$s) with any technique will essentially be unaffected by the satellite trails. Similarly, observations in the thermal IR regimes will be unaffected by the thermal emission of the satellites.
    \item Medium-duration exposures (100~s) with traditional fields of view are affected at a very low level during the astronomical night. Up to 0.5\% of imaging observations would be ruined during the twilights.
    \item Long exposures (1000s) with long-slit spectrographs: 0.3 to 0.4\% of the exposures would be ruined during the beginning and end of night, and up to 3\% of the exposures taken during twilight would be rendered useless. Short-slit and fibre-fed instruments are less affected. 
    \item Wide-field imaging and spectroscopic surveys: 1--5\% of the exposures would be ruined during the beginning and end of night, and at a higher level during twilight.
    \item Very wide-field imaging observations on large telescopes (such as those of the Vera C. Rubin Observatory), for which saturation and ghosting caused by a satellite will ruin the full exposure, would be severely affected: about 30\% of the exposures could be ruined at the beginning and end of the night. The situation is even worse during twilight (about 50\% of ruined images during astronomical twilight). Rubin observatory published a dedicated report based on an independent study (with different assumptions) indicating ``a 40\% impact on twilight observing time'' \citep{tyson20}. Only nights in  the middle of winter  would be completely unaffected. 
\end{itemize}

\B{This paper provides a first quantitative estimate of low-orbit satellite constellations on visible, NIR and thermal-IR astronomical observations, showing the key areas where follow-up assessments are needed and where collaborative efforts between the astronomy community, industry, and governments should focus. The results suggest that large telescopes like ESO's VLT and upcoming ELT will only be moderately affected, although some science cases may require the implementation of mitigation measures, such as scheduling of the observations or interruption of the exposures to allow a satellite cross the field of view. These mitigation measures have limitations, in particular for large fields of view. Wide-field surveys, in particular on large telescopes like the Vera Rubin Observatory, will be severely affected. Given the noted effect on wide-field surveys presented in this paper, further studies should examine the scientific implications on time-domain astronomy in general, asteroid and comet discovery and observation, planetary defence, and other affected science cases. }

\begin{acknowledgements}
We are very grateful to David Galad\'\i-Enr\`\i quez and Patrick Seitzer for their comments, assistance (including corrections!) and discussion, to Thomas Mueller for providing the thermal IR flux estimates, to Gie Han Tan for is assistance for the ALMA estimates and useful discussions. We are also grateful to Jo Andersen and Jason Spyromilio for useful comments on the manuscript. Finally, many thanks to our guardian angel, the anonymous referee who provided many insightful comments, validated many results, and spotted some errors and inconsistencies in the original manuscript.
\end{acknowledgements}

\bibliographystyle{aa} 
\bibliography{SatConst}

\end{document}